\documentclass[twocolumn,prb,showpacs,amsmath,amssymb]{revtex4}

\usepackage{graphicx}
\usepackage{subfigure}

\def\be{\begin{equation}}
\def\ee{\end{equation}}
\def\ba{\begin{eqnarray}}
\def\ea{\end{eqnarray}}

\begin{document}
			
\title{Downward Shift of Infrared Conductivity Spectral Weight at the DDW Transition: Role of
Anisotropy}

\author{Rouzbeh Gerami and Chetan Nayak}
\affiliation{Department of Physics, University of California,  
Los Angeles, California 90095--1547}
\date{\today}

\begin{abstract}
We consider the motion of conductivity spectral weight at a finite-temperature
phase transition at which $d_{x^2-y^2}$ density-wave (DDW) order develops.
We show that there is a shift of spectral weight to higher frequencies if
the quasiparticle lifetime is assumed to be isotropic, but a shift to lower
frequencies if the quasiparticle lifetime is assumed to be anisotropic.
We suggest that this is consistent with recent experiments on the
pseudogap phase of the cuprate superconductors and, therefore, conclude
that the observation of a downward shift in the spectral weight at the pseudogap
temperature does not militate against the DDW theory of the pseudogap.
\end{abstract}


\maketitle

\section {Introduction}

It has been proposed \cite{DDW} that the anomalous behaviors observed
in the pseudogap phase of the underdoped cuprates can be explained by assuming
the existence of a new long-range ordered density-wave
with $d_{x^2-y^2}$ symmetry (DDW) 
\cite {Chetan1} which competes with superconducting order.
As a result of this order, the excitation spectrum 
acquires a partial gap with the same symmetry and structure as the
gap observed in the pseudogap state. The DDW theory 
has been applied to explain several observed anomalies in the pseudogap
phase of the cuprates
\cite{DDWARPES, DDWHallAngle, DDWSTM, DDWSuperfluid,DDWHallNumber, Benfatto1, Benfatto2}.  Attempts
to directly observe it in neutron scattering are encouraging \cite{Mook}.

One of the arguments against the DDW theory is that it is
inconsistent with in-plane optical conductivity measurements.
The assumption has been that as the temperature is reduced below the
pseudogap transition temperature $T^*$, and a (partial) gap opens, some of the
low-energy excitations are lost and this should cause a reduction
in the low-frequency optical conductivity, which is compensated by
an increase in the spectral weight at higher frequencies.
Such behavior is seen in some charge density-wave systems \cite{CDW}, but
is not observed in the cuprates \cite{PuchkovJPC,TimuskRPP,BontempsPRL},
which might be taken as an indication that
the pseudogap is associated with superconducting fluctuations, which
would cause a downward motion of spectral weight (which would eventually
coalesce into a zero-frequency $\delta$-function in the superconducting state).

These experiments find no changes in the in-plane spectral weight
within their error bars \cite{PuchkovJPC,TimuskRPP,BontempsPRL}:
spectral weight lost at the low-energy end of the measured frequency range
appears (perhaps unexpectedly) at very low frequencies, and hence accompanies
a narrowing of the low-energy optical conductivity peak.
This phenomenon was interpreted in ref. \onlinecite{BontempsPRL}
as the lack of any in-plane optical evidence 
for the pseudogap. 

In this paper we show how the DDW theory of the pseudogap
can be consistent with this observed effect.
The key to this phenomenon, we believe, (as 
conjectured in \onlinecite{BontempsPRL}) is the 
strong anisotropy in the quasiparticle spectrum and scattering rate
in the cuprates. ARPES experiments clearly show 
that quasiparticles in the antinodal region of the Brillouin zone
($(k_x,k_y) \sim (0,\pm \pi), (\pm \pi, 0)$), where the gap opens in the
pseudogap phase, scatter much more strongly than they do in the nodal region 
($(k_x,k_y) \sim (\pm \pi, \pm \pi) $), where the gap is zero. (While the transport
lifetime is not the same as the quasiparticle lifetime observed in ARPES, we
assume that it has a similar anisotropy, as in ``hot spot'' and ``cold spot'' theories \cite{hotspots,coldspots}.)
The lost carriers, therefore, give a relatively small contribution to the conductivity.
The loss of these carriers can be cancelled by a
general reduction of all the scattering rates, as the temperature is reduced. 
As pointed out in ref.  \onlinecite{Basov02}, this may also explain
the observed behavior in several classes of spin-density-wave (SDW) and CDW
systems.

These ideas are borne out by a model calculation with the mean-field
DDW Hamiltonian and an {\it ansatz} for the quasiparticle lifetime.
We show that when the lifetime has an anisotropic
form motivated by ARPES measurements, the conductivity spectral weight shifts downward,
as in the cuprates. However, for an isotropic lifetime, the spectral weight shifts upward.

\section{Mean-Field Hamiltonian}

The mean-field Hamiltonian for the DDW state is 
\be
\label{eqn:mean-field-Ham}
    H_{\rm  DDW} = \sum_{{\bf k},\sigma}
    [( \epsilon_{\bf k}-\mu )  c^\dagger_{{\bf k}  \sigma}   c_{{\bf k}
    \sigma} +
    (i\Delta_{\bf k}  c^\dagger_{{\bf k} \sigma}c_{{\bf k+Q} \sigma} + \text{ h.c.})] 
\ee
where $c_{\bf k}$ is the annihilation operator of an electron in a state with momentum ${\bf k}$ and 
spin $\sigma$, $\mu$ is the chemical potential, 
$\Delta_{\bf k} = \langle c^\dagger_{\bf{k} \sigma} c_{\bf {k+Q} \sigma} \rangle  = \Delta_0 (\cos k_x - \cos k_y) /2 $ 
is the DDW order parameter and  ${\bf Q} = (\pi,\pi)$ is the DDW ordering wave vector, with
lattice spacing set to unity. The tight-binding band structure is given by 
$\epsilon_{{\bf k}} = \epsilon_{1 {\bf k}} + \epsilon_{2 {\bf k}}$, with

\be
    \epsilon_{1{\bf k}} = -2t (\cos k_x + \cos k_y)\:,  \hskip 0.4 cm
    \epsilon_{2{\bf k}} = 4t' \cos k_x  \cos k_y.
\ee
where $t$ and $t'$ are nearest-neighbor and next-neighbor hopping parameters. 

Introducing the two-component field operator
$\chi^\dagger_{\bf k \sigma} = \left( 
                      \begin {array} {cc} 
					    c^\dagger_{\bf k \sigma} & -i c^\dagger_{\bf k+Q \sigma} 				
			          \end{array}
			          \right)
$   
the mean field Hamiltonian can be written as
\be
    H_{DDW} = \sum_{{\bf k}, \sigma} \chi^\dagger_{\bf k \sigma} B_{\bf k} \chi_{\bf k \sigma}
\ee
with
\be
    B_{\bf k} = \left(
					 \begin{array}{ccc}
					    \epsilon_{\bf k}-\mu  & \Delta_{\bf k} \\
						\Delta_{\bf k}        & \epsilon_{\bf k+Q}-\mu
					 \end{array}
				\right) 
\ee
or
\be
	B_{\bf k} = (\epsilon_{2 \bf k} - \mu) + \epsilon_{1 \bf k} \sigma^3 + \Delta_{\bf k} \sigma^1
\ee
where $\sigma^i$ are the Pauli matrices and the sum is over half the original 
Brillouin zone (reduced Brillouin zone - RBZ), {\em i.e.} $|k_x| + |k_y| \leq \pi$, 
.

\begin{figure}
\centerline{\includegraphics[scale=0.75]{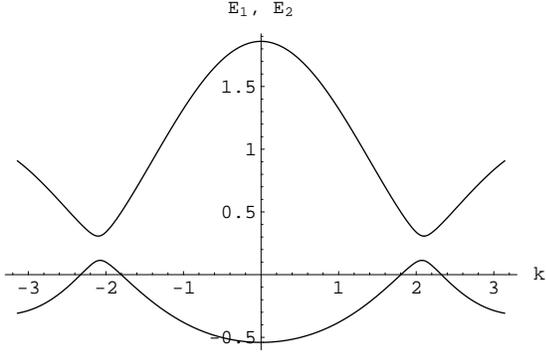}}
\caption{ DDW quasiparticle energy bands along the line $k_y = k_x/2$ in the Brillouin zone for $t=0.3, t'=0.09$,$\Delta_0=0.2$ and $\mu =-0.3$ eV. }
\label{Fig:EnSpect}
\end{figure}

Diagonalizing the Hamiltonian will then give the DDW quasiparticle energy bands 
\be
	E_{1,2}= (\epsilon_{2 \bf k} - \mu) \pm \sqrt{\epsilon^2_{1 \bf k} + \Delta^2_{\bf k}}
\ee
as depicted in figure \ref{Fig:EnSpect}.

\section{Green Function}

The "non-interacting" Nambu Green's function can then be obtained by
inverting the matrix $B_{\bf k}$ . (For now, the only relevant
interactions are the electron-electron interactions which generate the DDW coupling.
All other interactions including  quasiparticle-quasiparticle and quasiparticle-impurity
interactions will be taken into account later by assuming a non-zero self energy).

The $2 \times 2$ Nambu Green function is defined by   
\begin{multline}
G_0({\bf k},t) = \langle T \chi_{\bf k}(t) \chi^\dagger_{\bf k}(0)\rangle \\
	  =\left(
	  \begin{array}{ccc}
	    \langle T c_{\bf k  }(t) c^\dagger_{\bf k}(0)\rangle & i\langle T c_{\bf k  }(t) c^\dagger_{\bf k+Q}(0)\rangle \\ 
	  -i\langle T c_{\bf k+Q}(t) c^\dagger_{\bf k}(0)\rangle &  \langle T c_{\bf k+Q}(t) c^\dagger_{\bf k+Q}(0)\rangle \\ 
      \end{array}
	  \right) 
\end{multline}

The Fourier transform of the Green function matrix is then
\begin{multline}
G_0({\bf k},\omega) = \int dt   e^{i \omega t} G_0({\bf k}, t) \\
= \frac{1}{(\omega + i\delta)-(\epsilon_{2 \bf k} - \mu) - \epsilon_{1 \bf k} \sigma^3 - \Delta_{\bf k} \sigma^1} \\
= \frac {(\omega - (\epsilon_{2 \bf k} - \mu) + \epsilon_{1 \bf k} \sigma^3 + \Delta_{\bf k} \sigma^1} 
        {\omega -(\epsilon_{2 \bf k} - \mu)+ i\delta)^2 - \epsilon_{1 \bf k}^2 + \Delta_{\bf k}^2}
\end{multline}

We now consider the effects of impurity scattering and `residual' electron-electron interactions. Here,
we will capture their combined effect by introducing a non-zero single-particle self-energy 
$\Sigma ({\bf k}, \omega) = \Sigma_1 ({\bf k}, \omega) + i \Sigma_2 ({\bf k}, \omega) $,
where the real and imaginary parts give the energy renormalization and quasiparticle lifetime,
respectively. Neglecting the shift in the excitation energy due to the real part
$\Sigma_1 ({\bf k}, \omega)$, the self-energy can be written in terms 
of the quasiparticle lifetime as  
\be
\Sigma ({\bf k},\omega) = -\frac{i}{\tau ({\bf k},\omega)}  
\ee
where $1/\tau ({\bf k},\omega)$ is the quasiparicle scattering rate (inverse lifetime). 
Including the self-energy, the full Green's function $G({\bf k}, \omega)$ will then be given according
to Dyson's equation by
\be
 G({\bf k},\omega) =  \left(G_0({\bf k},\omega)^{-1} - \Sigma({\bf k},\omega)\right)^{-1} 
\ee

The spectral function $A({\bf k},\omega)$ can then be calculated by taking the imaginary part
of the Green function 
\begin{widetext}
\be
A({\bf k},\omega)= -\frac{1}{\pi} \text{Im  } G({\bf k}, \omega)
 = \frac{1}{\pi \tau} 
  \frac{(\omega -\epsilon_{2 \bf k} + \mu)^2  + (\epsilon_{1 \bf k}^2 + \Delta_{\bf k}^2)+(1/\tau)^2 
   + 2(\omega -\epsilon_{2 \bf k} + \mu)(\epsilon_{1 \bf k} \sigma^3+ \Delta_{\bf k} \sigma^1)}
 {[(\omega - E_{1 \bf k}) (\omega - E_{2 \bf k}) - (1/\tau)^2]^2 + (2(\omega -\epsilon_{2 \bf k} + \mu) / \tau)^2} 
\ee
\end{widetext}

\begin{figure}
\center
\centerline{\includegraphics[scale=0.65]{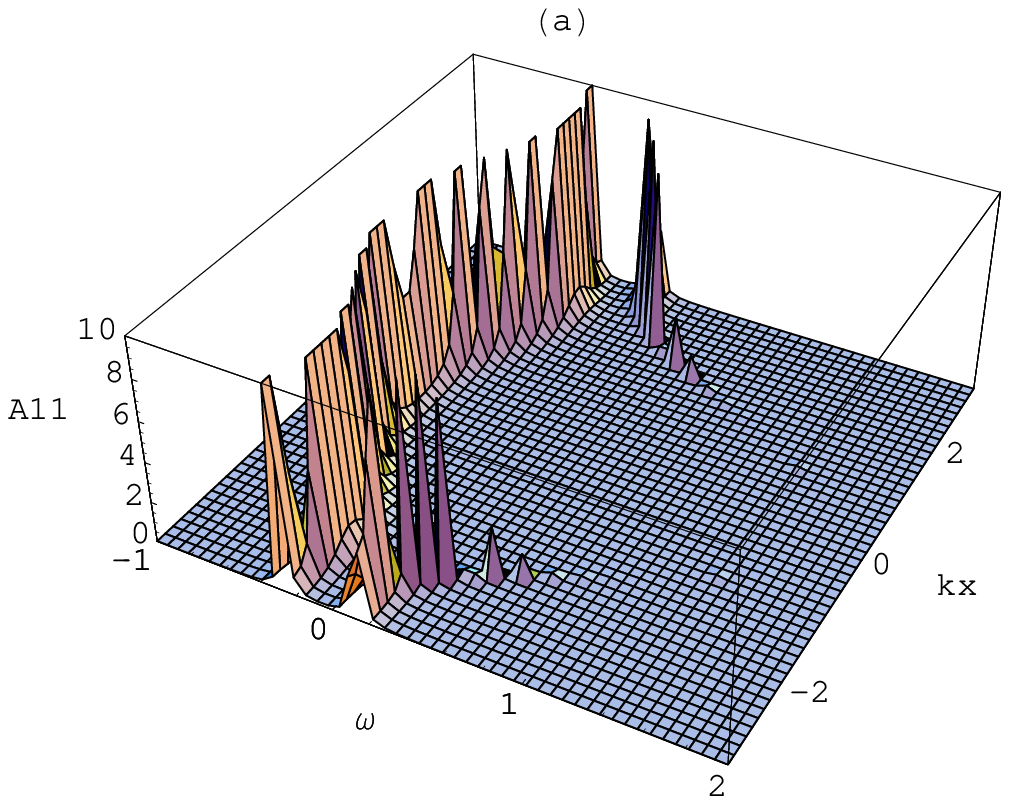}}
\centerline{\includegraphics[scale=0.65]{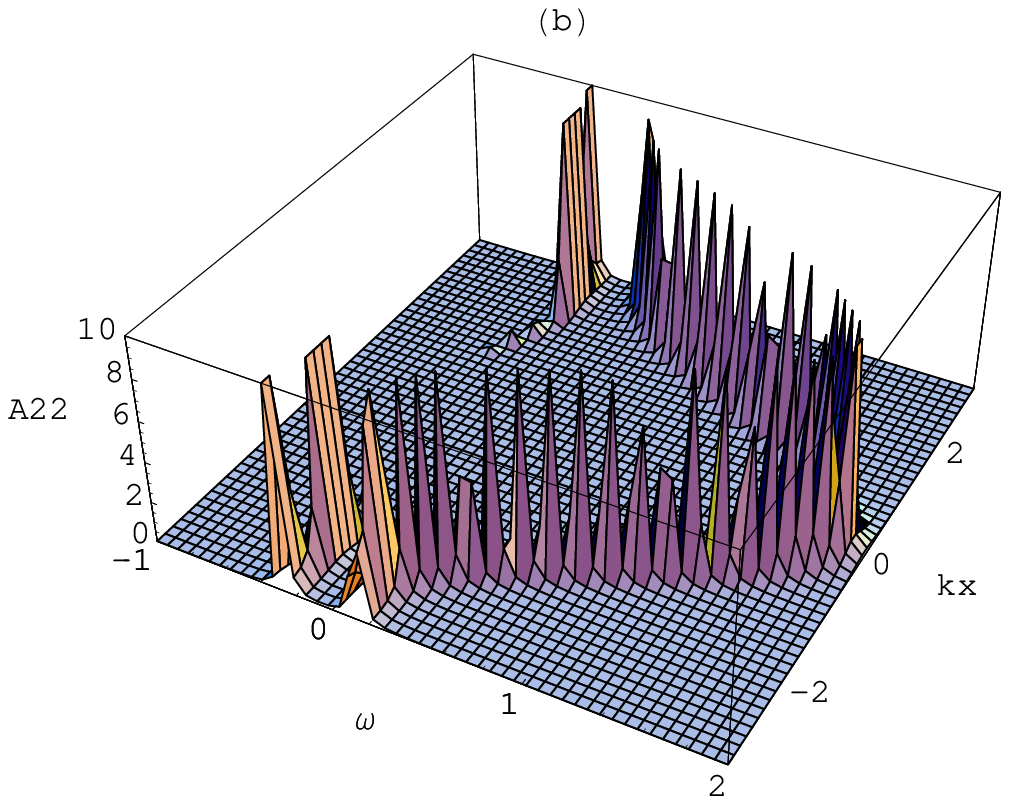}}
\caption{Elements of the DDW spectral function matrix for $1/\tau=0.005$ eV: (a)$A_{11}(\omega, k_x, k_y=0)$ and (b)$A_{22}(\omega, k_x, k_y=0)$. 
Other parameters are the same as in figure \ref{Fig:EnSpect}.}
\label{Fig:A}
\end{figure}

In Figure \ref{Fig:A}, the diagonal elements of the Nambu spectral
function matrix $A(\omega, k_x, k_y=0)$ are plotted against $k_x$ and $\omega$.
For demonstration purposes, the scattering rate here is assumed to be a constant 
$1/\tau = 0.02$ eV (independent of momentum and frequency).
The diagonal entries $A_{11}$ and $A_{22}$ have peaks centered 
mostly at energies corresponding to the upper and lower energy bands ($E_1$ and $E_2$) respectively.

\section {Optical conductivity}

Having found the spectral function $A(\omega, \bf{k})$, the real part of the
AC conductivity can now be calculated by using the Kubo formula 
\be
\text{Re  } \sigma_{xx}(\omega) = \frac{1}{\omega} \text{Im  } \Pi_{xx}(i \omega_n \rightarrow \omega + i \delta ) 
\ee
where $\Pi(i \omega_n)$ is the Fourier tansform of the current-current
correlation function in Matsubara formalism 
\be
\Pi_{xx} ( i \omega_n) =  \int_0^\beta d\tau e^{i \omega_n \tau} \Pi_{xx} ( \tau) 
\ee
with
\be
\Pi_{xx} ( \tau) = \langle T_\tau j_x ( \tau) j_x (0) \rangle
\ee

The current operator for the DDW quasiparticles can be obtained by minimally-coupling the
mean-field Hamiltonian (\ref{eqn:mean-field-Ham}) to the electromagnetic field, ${\bf A}$,
and differentiating once with respect to ${\bf A}$:
\begin{multline}
{\bf j} =  \sum _{RBZ} \biggl[{{\bf v}_{F2}}({\bf k}) \left(\chi_{\bf k}^\dagger \chi_{\bf k}\right)
+ {{\bf v}_{F1}}  ({\bf k})\left(\chi_{\bf k}^\dagger \sigma^3 \chi_{\bf k} \right)\\
+ {{\bf v}_\Delta}({\bf k})\left(\chi_{\bf k}^\dagger \sigma^1 \chi_{\bf k} \right)\biggr]
\end{multline}  
where ${{\bf v}_{F1}}({\bf k}) = {\bf \nabla_k} \epsilon_1({\bf k})$ 
,${{\bf v}_{F2}}({\bf k}) = {\bf \nabla_k} \epsilon_2({\bf k})$ and 
 ${{\bf v}_\Delta}({\bf k})={\bf \nabla_k}\Delta({\bf k})$.

Using this form of the current operator, the current-current correlation function can be written
in terms of the elements of the imaginary-time Nambu Green's function ${\cal G}_{ij}$ (the non-interacting
Green's function $ G_{ij}$ is now promoted to the interacting one ${\cal G}_{ij}$, to take the effect of the 
residual interactions into account). We will have  
\begin{multline}
\langle T_\tau j(\tau) j(0) \rangle = -\sum _{RBZ}\bigl(  
    {[{{\bf v}_{F2}}({\bf k})]^2}   {\rm tr}( {\cal G} (-\tau) {\cal G} (\tau))\\  
 +	{[{{\bf v}_{F1}}({\bf k})]^2}   {\rm tr}({\sigma^3} {\cal G} (-\tau) {\sigma^3}{\cal G} (\tau))\\  
 +  {[{{\bf v}_\Delta}({\bf k})]^2} {\rm tr}\left({\sigma^1}{\cal G}(-\tau) {\sigma^1} {\cal G}(\tau)\right)\bigr)\\
 +  ({{\bf v}_{F1}}.{{\bf v}_{F2}}) {\rm tr}\left( ({\sigma^3}{\cal G}(-\tau) + {\cal G}(-\tau){\sigma^3}) {\cal G}(\tau) \right)\bigr)\\
 +  ({{\bf v}_{F2}}.{{\bf v}_{\Delta}}) {\rm tr}\left( ({\sigma^1}{\cal G}(-\tau) + {\cal G}(-\tau){\sigma^1}) {\cal G}(\tau) \right)\bigr)\\
 +  ({{\bf v}_{F1}}.{{\bf v}_{\Delta}}) {\rm tr}\left( ({\sigma^1}{\cal G}(-\tau){\sigma^3} + {\sigma^3}{\cal G}(-\tau){\sigma^1}) {\cal G}(\tau) 
\right)\bigr)\\
\end{multline}
In this equation, we have ignored vertex corrections. These are important when the scattering
rate is strongly angle-dependent, as they distinguish the transport and quasiparticle lifetimes
(for instance, through a $(1-\cos\theta)$ factor) and also distinguishing umklapp scattering
from momentum-conserving scattering. In what follows, we will assume that the replacement
$\tau\rightarrow\tau_{\rm tr}$ is made ({\em i.e.} ignore vertex corrections). Reference\cite {vertex} has 
considered the vertex correction for DDW conductivity. 

Writing ${\cal G}$ in terms of the spectral function  $A(\bf k, \omega)$,
evaluating the Matsubara sum, and doing the 
analytic continuation $(i \omega_n \rightarrow \omega + i\delta)$,
we find the optical conductivity to be \cite {Mahan}

\begin{figure}
\centerline{\includegraphics[scale=0.80]{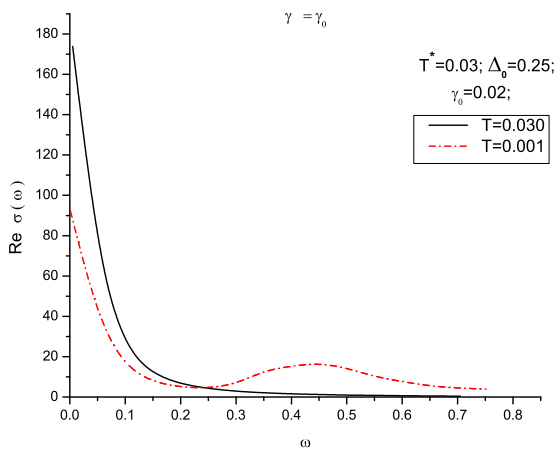}}
\caption{ Real part of the optical conductivity for a constant scattering rate $\gamma=\gamma_0 = 0.02$, $\Delta_0 = 0.25$ and $T^*=0.030$. }
\label{Fig:BigGap}
\end{figure}

\begin{widetext}
\begin{multline}
\sigma(\omega)\sim 
\frac{1}{\omega} \sum_{RBZ} \int_{-\infty} ^\infty \frac{d\varepsilon}{2\pi}
\left[n_F(\varepsilon)-n_F(\varepsilon+\omega) \right]\times\\
\biggl\{  {\left[{{\bf v}_{F2}}({\bf k})\right]^2} \Bigl[   A_{11}(k,\varepsilon) A_{11}(k,\varepsilon+\omega)  
                                                       +2 A_{12}(k,\varepsilon) A_{12}(k,\varepsilon+\omega)
                                                       +  A_{22}(k,\varepsilon) A_{22}(k,\varepsilon+\omega) \Bigr]  \\
        + {\left[{{\bf v}_{F1}}({\bf k})\right]^2} \Bigl[  A_{11}(k,\varepsilon) A_{11}(k,\varepsilon+\omega)  
                                                       -2 A_{12}(k,\varepsilon) A_{12}(k,\varepsilon+\omega)
                                                       +  A_{22}(k,\varepsilon) A_{22}(k,\varepsilon+\omega) \Bigr]  \\
   + {\left[{{\bf v}_{\Delta}}({\bf k})\right]^2} \Bigl[  A_{22}(k,\varepsilon) A_{11}(k,\varepsilon+\omega)  
                                                       +2 A_{12}(k,\varepsilon) A_{12}(k,\varepsilon+\omega)
                                                       +  A_{11}(k,\varepsilon) A_{22}(k,\varepsilon+\omega) \Bigr]  \\
+2 {{\bf v}_{F1}}({\bf k}). {{\bf v}_{F2}}({\bf k})   \Bigl[A_{11}(k,\varepsilon) A_{11}(k,\varepsilon+\omega) 
                                                        - A_{22}(k,\varepsilon) A_{22}(k,\varepsilon+\omega) \Bigr]  \\
+2 {{\bf v}_{F2}}({\bf k}). {{\bf v}_\Delta}({\bf k}) \Bigl[A_{12}(k,\varepsilon) ( A_{11}(k,\varepsilon+\omega)+ A_{22}(k,\varepsilon+\omega)) 
                                                       +  ( A_{11}(k,\varepsilon)+A_{22}(k,\varepsilon)) A_{12}(k,\varepsilon+\omega)  \Bigr]\\
+2 {{\bf v}_{F1}}({\bf k}). {{\bf v}_\Delta}({\bf k}) \Bigl[A_{12}(k,\varepsilon) ( A_{11}(k,\varepsilon+\omega)- A_{22}(k,\varepsilon+\omega)) 
                                                       +  ( A_{11}(k,\varepsilon)- A_{22}(k,\varepsilon)) A_{12}(k,\varepsilon+\omega)  \Bigr]
 \biggr\}  
\end{multline}
\end{widetext}
where ${n_F}(\epsilon)$ is the Fermi distribution function. For demonstration purposes, in
fig.~\ref{Fig:BigGap} the 
real part of the optical conductivity has been plotted against $\omega$
for two different temperatures (one above $T^*$, depicted with a solid line,
one below $T^*$, depicted with a dashed-dotted line), assuming that
the quasiparticle lifetime is a constant
(temperature and momentum independent) and the gap is unrealistically big
($W_0 = 0.25$ eV). As expected an upward shift of the
SW occurs when the gap opens. Similar calculations have also been done in \cite {Valenzuela}.

\section {Quasiparticle lifetime}

Applying the formulas of the preceding section to the underdoped cuprates
presupposes that the quasiparticle picture makes sense there. This
is questionable, particularly in the anti-nodal regions, where even lowest-order
perturbation theory around the DDW mean-field Hamiltonian \cite{DDWARPES}
predicts short lifetimes which may indicate a breakdown of quasiparticles.
However, we will compute the conductivity in the quasiparticle approximation
to show that, even at this level, an upward shift of spectral weight is not expected.
If there are no quasiparticles at the anti-nodes, then the situation may be
even better.

In order to proceed with this strategy, we need one final ingredient,
an {\it ansatz} for the quasiparticle scattering
rate ${1}/{\tau({\bf k},\omega; T)}= \gamma({\bf k},\omega; T)$, as a function of 
momentum, frequency, and temperature in the underdoped cuprates.
A number of angle-resolved photoemission experiments
have measured the inverse lifetime (imaginary part of the self energy), as a function of these different
parameters. In these experiments, the width of the quasiparticle peaks in the
energy distribution curves (EDC's) or momentum 
distribution curve (MDC's) are measured as functions of momentum, energy and temperature
\cite{VallaPRL00, VallaScience99, Kaminski0404385,KaminskiPRL00}.
While the transport lifetimes are not necessarily identical to the quasiparticle lifetimes
(or, equivalently, the vertex corrections are not necessarily small), we expect that they
will have a similar anisotropic behavior. For the purposes of this model calculation,
we will simply take them to be the same. We emphasize that the lifetimes which
we adopt below are for illustrative purposes since our main goal is to show that
an upward movement of spectral weight is not a necessary concomitant of
the emergence of DDW order at finite temperature. We are not making any claims here
about the correctness of these lifetimes.

We take the imaginary part
of the self-energy (quasiparticle scattering rate) to have the form
\be
\label{eqn:lifetime-split}
\Sigma_2 ({\bf k},\omega; T) =  \Sigma_2(\omega; T) + \Gamma ({\bf k}).
\ee
where $\Sigma_2(\omega; T)$ is temperature and frequency dependent
with no momentum dependence and $\Gamma ({\bf k})$ is strongly momentum dependent. 

Such a form has been motivated by marginal Fermi liquid phenomenology \cite{MFL,PNAS} (MFL).
In this way of analyzing the data, it is assumed that the quasiparticle lifetime which comes
from electron-electron scattering is independent of temperature and linear in energy for
small temperatures and linear in temperature, 
independent of the binding energy for higher temperatures: 
\be
\Sigma^{MFL}_2(\omega; T) = \lambda \text{Max}(|\omega|, T).
\ee
In this {\it ansatz}, all of the angular dependence comes from {\it elastic} electron-electron scattering.
This is a convenient form, but we will show that our results hold even for
some others.

For instance, we repeat our calculations with the standard Fermi liquid (FL) quasiparticle lifetime:
\be
\Sigma^{FL}_2(\omega; T) = \lambda  \text{Max}(\omega^2, T^2).
\ee
again assuming that the angular dependence comes from $\Gamma ({\bf k})$. In order to show that 
these particular forms of $\omega$-dependence are not playing a big role in shifts of SW, we have also 
tried $\omega$-independent forms of these scattering rates: $\Sigma^{T-Linear}_2(\omega; T) = \lambda  T$ and
$\Sigma^{T^2}_2(\omega; T) = \lambda  T^2$.  

While this form of the quasiparticle lifetime does not give the correct DC conductivity,
it is still useful as a check because our goal is to emphasize the role of anisotropy
and show that it can lead to a downward shift of spectral weight, regardless of
its detailed frequency and temperature dependence.

The quasiparticle lifetime is strongly anisotropic \cite{Kaminski0404385}:
excitations in the  antinodal region are more strongly scattered than the
ones in the nodal region by up to a factor of 5.
Hence, the assumed form (\ref{eqn:lifetime-split}) necessitates that $\Gamma ({\bf k})$
be a strongly anisotropic function of $\bf k$. We take the form:
\be
\Gamma_{\rm Aniso} ({\bf k}) = \gamma_0 (1+ (\cos k_x - \cos k_y)^2),
\ee
where $\gamma_0$ is the scattering strength in the nodal region.

In order to check the role of anisotropy in the eventual shifting of the SW, 
we also check the case in which $\Gamma ({\bf k})$ is (unrealistically) isotropic:  
\be
\Gamma_{\rm Iso} ({\bf k}) = \gamma_0 .
\ee

\section {Results}

\begin{figure}
\begin{center}
\centerline{\includegraphics[scale=0.90]{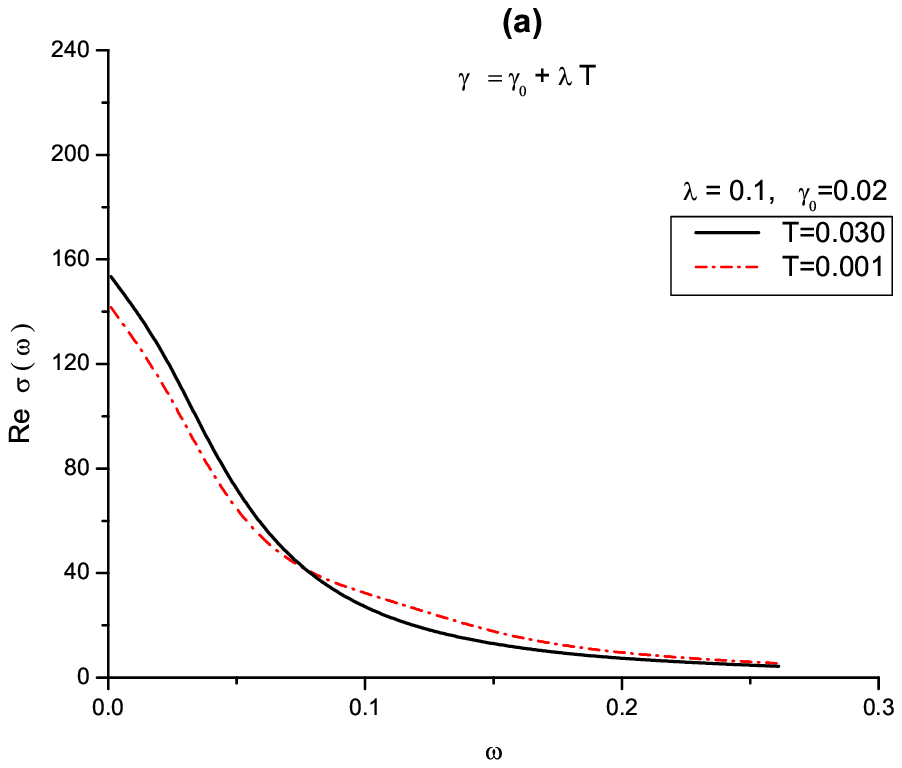}}
\centerline{\includegraphics[scale=0.90]{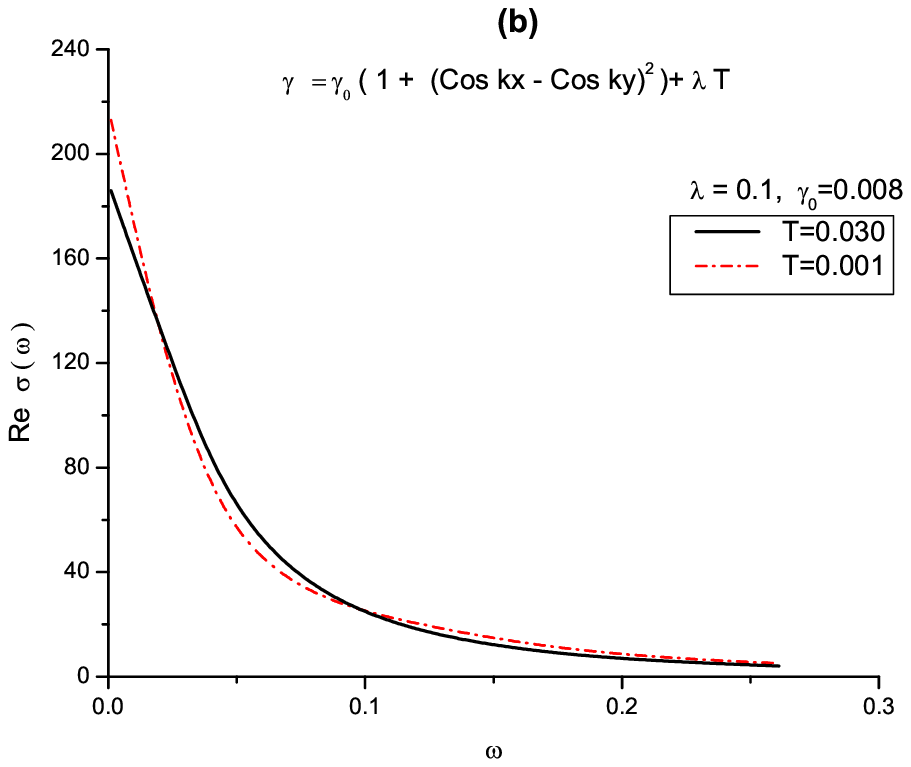}}
\caption{ Real part of the optical conductivity for (a) isotropic scattering rate
with $\gamma_0=0.020$ and  (b) anisotropic scattering rate with $\gamma_0=0.008$.
Quasiparticle lifetime is taken to be linear in temperature and 
independent of $\omega$.}
\label{Fig:Linear}
\end{center}
\end{figure}

\begin{figure}
\begin{center}
\centerline{\includegraphics[scale=0.90]{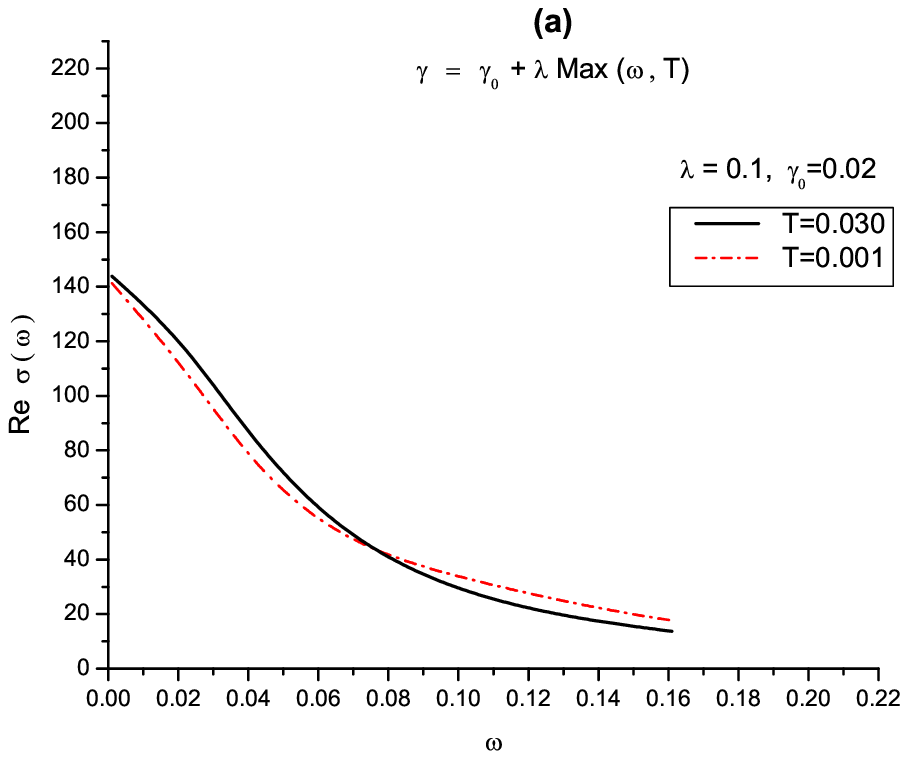}}
\centerline{\includegraphics[scale=0.90]{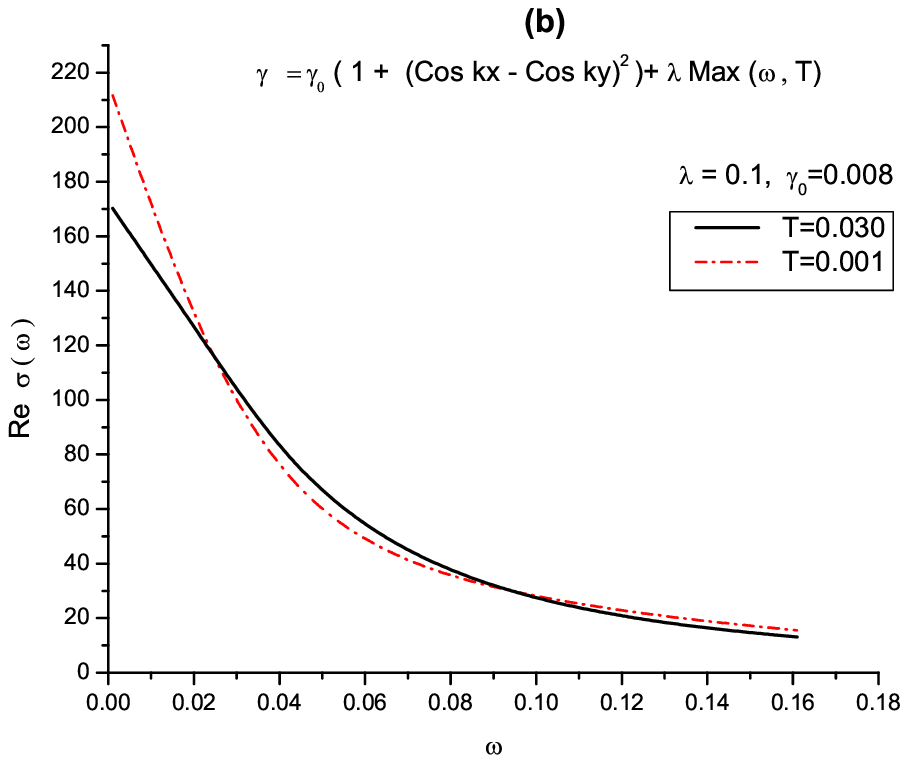}}
\caption{Real part of the optical conductivity for (a) isotropic scattering rate with $\gamma_0=0.020$ and 
(b) anisotropic scattering rate with $\gamma_0=0.008$. Quasiparticle lifetime's temperature and frequency dependence 
 is given by the marginal Fermi liquid theory.}
\label{Fig:MFL}
\end{center}
\end{figure}

\begin{figure}
\begin{center}
\centerline{\includegraphics[scale=0.90]{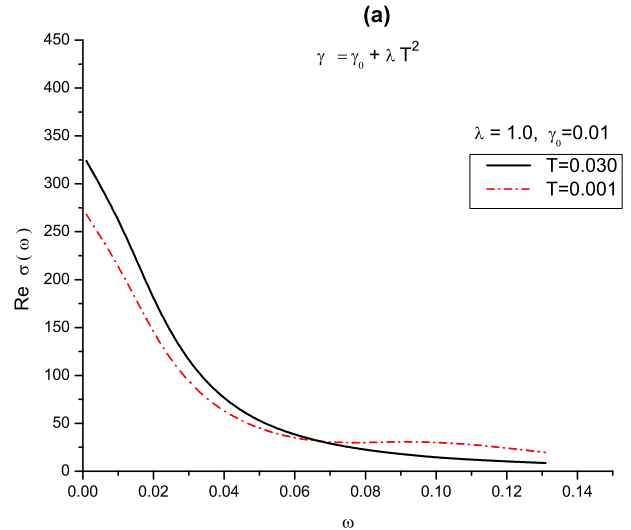}}
\centerline{\includegraphics[scale=0.90]{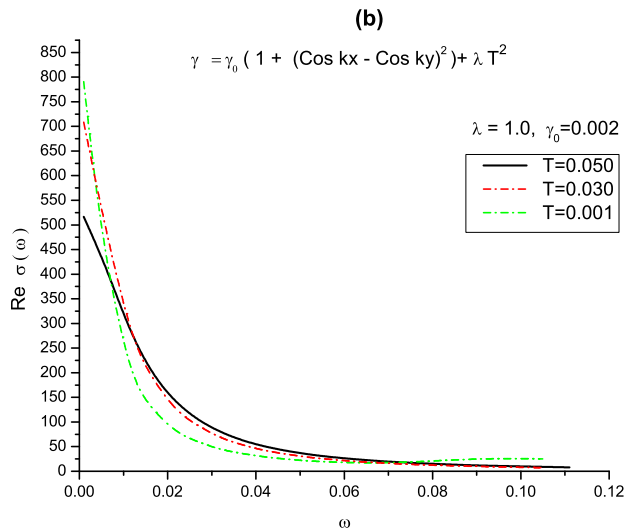}}
\caption{Real part of the optical conductivity for (a) isotropic scattering rate with $\gamma_0=0.010$ and 
(b) anisotropic scattering rate with $\gamma_0=0.002$. Quasiparticle lifetime is taken be to independent of $\omega$ have a $T^2$ temperature 
dependence.}
\label{Fig:T2}
\end{center}
\end{figure}

\begin{figure}
\begin{center}
\centerline{\includegraphics[scale=0.90]{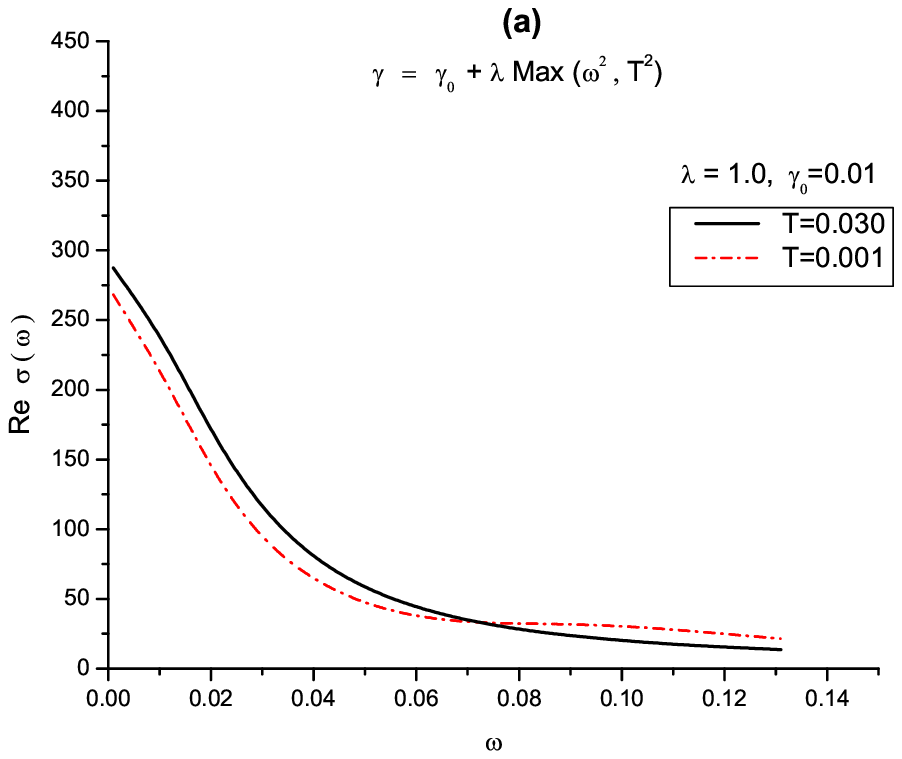}}
\centerline{\includegraphics[scale=0.90]{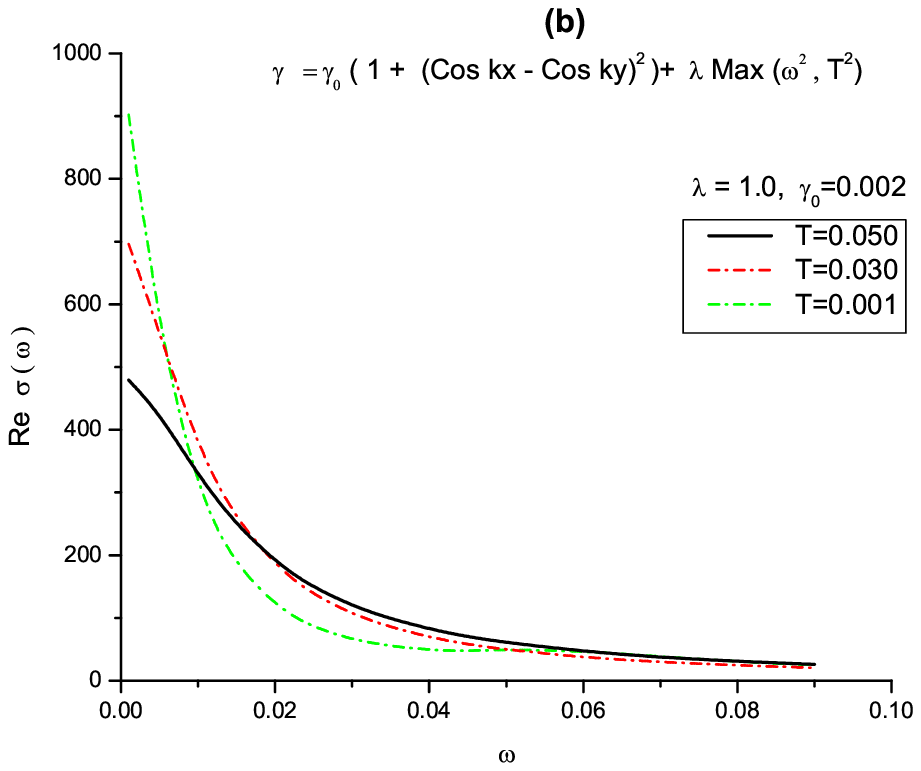}}
\caption{ Real part of the optical conductivity for (a) isotropic scattering rate with $\gamma_0=0.020$ and 
(b) anisotropic scattering rate with $\gamma_0=0.008$. Quasiparticle lifetime's temperature and frequency dependence 
 is given by the Fermi liquid theory.}
 \label{Fig:FL}
\end{center}
\end{figure}

In Figures \ref{Fig:Linear} to \ref{Fig:FL}, we have plotted the real part of the optical conductivity vs $\omega$, when the  momentum-independent 
part of the lifetime is given by the four different forms listed above. In each figure the isotropic case is compared with the anisotropic one with 
the same set of parameters (listed in the figure captions) except for $\gamma_{0}$, which is smaller in the 
anisotropic case because of its extra (larger than 1) prefactor. In all cases we have $t=0.3$, $t'=0.09$,
$\mu = -0.3$ and $T^*= 0.03$.

It is clear that in figures 4a, 5a, 6a, 7a, in which the scattering rate is isotropic,
there is an upward movement of spectral weight (though it is small in some cases).
However, in figures 4b, 5b, 6b, 7b, in which the scattering rate is anisotropic,
there is a clear downward movement of spectral weight. (Broadening the Drude peak has 
also been seen in \cite{Valenzuela} and \cite{vertex}, in which scattering has been isotropic.)

\section {Conclusions}

Recent data by Santander-Syro {\em et al} \cite {BontempsPRL},
has questioned the previous belief that the opening of the pseudogap 
in the underdoped cuprates can be detected by in-plane optical conductivity
measurements. The unexpected result that as the 
temperature is reduced (and although a gap opens in the
antinodal region of the Brillouin zone), the spectral weight is still 
transferred to lower frequencies, was interpreted as lack of any pseudogap
signature in the optical data. In this  paper we showed that this effect is
consistent with the DDW theory of the pseudogap state of the 
underdoped cuprates. 

In the four sets of graphs of the previous section it can clearly be seen 
that the key factor in deciding which way the SW is transferred
is isotropy or anisotropy of the scattering rate. It is shown that
regardless of the form of the temperature and frequency 
dependence of the quasiparticle lifetime ({\em e.g.} Fermi liquid
or non-Fermi liquid) the SW is shifted upward for the isotropic
scattering rate while it is shifted downward for the anisotropic case.
(We have not tried to find a form for the quasiparticle lifetime which
correctly reproduces all of the transport data, but have focussed on
the role of anisotropy and have shown that similar results are obtained
for several different forms of frequency and temperature dependence.)

We believe the explanation to be as follows: the pseudogap opens
in the antinodal region where the carriers are more
strongly scattered than the ones in the nodal region, where there is no gap.
Therefore, we have lost those excitations which already
gave relatively little contribution to the transport properties of the normal state.
Furthermore, by reducing the temperature, the scattering rate of all the
excitations is reduced.
Our results clearly show that in the (more realistic) anisotropic case, the 
effect of lost excitations due to the gap opening can be more than
canceled by a temperature-dependent reduction of the scattering 
rate for the rest of the excitations and hence a downward shift of the
optical spectral weight. It is obvious that anisotropy is the key here, 
since for similar parameters for isotropic scattering rate, an upward transfer is observed.    
We note that similar considerations should apply to the case of
2H-TaSe$_2$\cite{Forro}.

\acknowledgements

We would like to thank Sudip Chakravarty and Dmitri Basov for discussions.
This work has been supported by the NSF under Grant No. DMR-0411800.


\end{document}